\documentclass{article}

\usepackage[english]{babel}

\usepackage[letterpaper,top=2cm,bottom=2cm,left=3cm,right=3cm,marginparwidth=1.75cm]{geometry}

\usepackage{amsmath,booktabs}
\usepackage{graphicx}
\usepackage{bbm}
\usepackage[colorlinks=true, allcolors=blue]{hyperref}
\usepackage{float}
\usepackage{subfigure}
\usepackage{lineno,hyperref}
\usepackage{amssymb} 
\usepackage[ruled,vlined]{algorithm2e}
\usepackage{mwe}
\usepackage{caption}
\usepackage{tabularx}
\usepackage[numbers]{natbib}
\usepackage{authblk}

\modulolinenumbers[5]

\newtheorem{theorem}{Theorem}[section]
\newtheorem{definition}[theorem]{Definition}

\newtheorem{example}[theorem]{Example}

\title{Optimal Entry and Exit with Signature in Statistical Arbitrage}
\author[1]{Boming Ning\thanks{Corresponding author; Email: \texttt{ningb@purdue.edu}}}
\author[2]{Prakash Chakraborty}
\author[1]{Kiseop Lee}
\affil[1]{Department of Statistics, Purdue University}
\affil[2]{Department of Industrial and Manufacturing Engineering, The Pennsylvania State University}

\providecommand{\keywords}[1]{\textbf{\textit{Keywords---}} #1}

\begin{document}
\maketitle

\begin{abstract}
In this paper, we explore an optimal timing strategy for the trading of price spreads exhibiting mean-reverting characteristics. A sequential optimal stopping framework is formulated to analyze the optimal timings for both entering and subsequently liquidating positions, all while considering the impact of transaction costs. Then we leverages a refined signature optimal stopping method to resolve this sequential optimal stopping problem, thereby unveiling the precise entry and exit timings that maximize gains. Our framework operates without any predefined assumptions regarding the dynamics of the underlying mean-reverting spreads, offering adaptability to diverse scenarios. Numerical results are provided to demonstrate its superior performance when comparing with conventional mean reversion trading rules.
\end{abstract}


\section{Introduction} 
\label{sec1}
Statistical arbitrage, also referred to mean reversion trading, is a trading strategy that employs statistical methods to identify pricing discrepancies between assets and capitalize on them for profit. It is widely used by traders and fund managers to simultaneously establish positions in two correlated assets and speculate on the resultant spread's path behavior. While modeling the dynamics of a single asset can be challenging, statistical models excel in capturing the mean-reverting behavior exhibited by pairs of assets or securities.
The mean reversion trading process comprises three fundamental steps: (1) the identification of co-moving assets, (2) the construction of spreads, and (3) the design of trading rules. The first two steps can be referred to as the formation period, while the last constitutes the trading period.

The process of identifying assets for statistical arbitrage relies on principles grounded in market dynamics, such as the co-movement of risk-on assets in the same direction or the correlation between stocks within a shared sector. For a trading pair to be considered profitable, it is imperative that the two assets exhibit a certain degree of correlation. This strategy entails the simultaneous purchase of one asset and sale of another, thereby achieving market neutrality within the designated group or sector, with the ultimate goal of capitalizing on the subsequent price convergence between these two assets. Empirical research has substantiated the presence of mean-reverting price behavior in various instances, including pairs of stocks and/or ETFs, disparities between futures and their corresponding spot prices, and spreads between physical commodities and commodity stocks/ETFs. Additionally, automated methodologies have been developed to identify portfolios with mean-reverting characteristics. In the context of this paper, we undertake a preliminary selection of several stock pairs in the US market, guided by human expertise, spanning diverse sectors, ranging from airlines to the technology industry. Each pair comprises two stocks originating from the same industry, thereby affording a comprehensive examination of mean reversion within different sectors.
  
The constructed spreads are essential to possess a mean-reverting characteristic, as this property underpins the opportunities available to traders engaged in statistical arbitrage. To illustrate, when the price of a spread deviates below its long-term mean, traders adopt a long position in the spread and await its reversion to the mean, thereby securing a profitable outcome. In the numerical experiments of the paper, we embrace a statistical approach to construct these spreads. For any given pair of stocks, we configure the spread to best conform to the Ornstein-Uhlenbeck (OU) model. More precisely, we ascertain the optimal ratio between the two stocks, as well as the model parameters, in a manner that maximizes the likelihood of the resulting spread time series. Notably, this method represents an extension of the conventional maximum likelihood estimation (MLE) approach, which traditionally focuses solely on determining the model parameters.

Within the trading period, investors confront the intricate task of deciding the optimal trading rules for both entering and exiting a position, a determination contingent upon the dynamic fluctuations in the price of the risky asset. Investors are presented with the choice of either entering the market immediately or exercising patience while monitoring the prevailing market prices, awaiting a more propitious moment. After executing the initial trade, the investor encounters the consequential decision of pinpointing the most advantageous instance for closing the position. This challenge motivates the central inquiry of our research—namely, the investigation of the optimal timing of trades.

In response to this challenge, we introduce a optimal trading timing framework designed to systematically assess the optimal moments for both initiating and subsequently liquidating positions. Our methodology formulates a sequential optimal stopping problem to analyze
the optimal trading timings, and capitalizes on the signature optimal stopping method, a powerful tool for solving this sequential optimal stopping problem, consequently revealing the precise timings for entry and exit that yield maximal gains. Importantly, our framework is adapted to a broad spectrum of mean reversion dynamics, with the Ornstein-Uhlenbeck (OU) model serving as a prime illustration in our experimental investigations.

The remainder of this paper is structured as follows. In Section~\ref{sec2}, we offer a comprehensive review of relevant literature. Section~\ref{sec3} lays the foundation by introducing the fundamental concepts pertaining to the signature. Within Section~\ref{sec4}, we expound upon the signature optimal stopping method. In Section~\ref{sec5}, we elaborate on our proposed framework for optimizing entry and exit decisions, presenting the associated solutions. The empirical validation of our methodology through numerical experiments, encompassing both simulated and real-world data, is detailed in Section~\ref{sec6}. Finally, Section~\ref{sec7} provides concluding remarks.

\section{Related Research} 
\label{sec2}
Examples of mean-reverting spreads are well-documented in a variety of empirical studies spanning diverse asset classes. They have been observed in pairs of stocks and ETFs (\citeauthor{gatev2006pairs} (\citeyear{gatev2006pairs}), \citeauthor{avellaneda2010statistical} (\citeyear{avellaneda2010statistical}), \citeauthor{Montana2011} (\citeyear{Montana2011}), \citeauthor{meanReversionBook} (\citeyear{meanReversionBook})), futures contracts (\citeauthor{futuresBS} (\citeyear{futuresBS}), \citeauthor{futuresDaiKwok} (\citeyear{futuresDaiKwok})), physical commodity and  commodity stocks/ETFs (\citeauthor{kanamura2010profit} (\citeyear{kanamura2010profit})), as well as cryptocurrencies (\citeauthor{leung2019constructing} (\citeyear{leung2019constructing})). 
 
The work of \citeauthor{gatev2006pairs} (\citeyear{gatev2006pairs}) stands as one of the pioneering studies in the realm of mean reversion trading, often referred to as pairs trading due to their method's reliance on pairs to construct a mean-reverting process. They introduce the widely adopted Distance Method (DM) and conduct empirical testing using CRSP stocks spanning from 1962 to 2002. The DM method initiates a trading position for a pair of assets when their prices deviate from each other by more than two historical standard deviations, subsequently closing the position when the prices converge again. Their findings reveal an excess return of 1.3 $\%$ for the top 5 pairs identified by the DM method and 1.4$\%$ for the top 20 pairs. Furthermore, \citeauthor{do2012pairs} (\citeyear{do2012pairs}) delve into the profitability of pairs trading while accounting for transaction costs. Their research extends the understanding of the practicality of pairs trading by incorporating the impact of transaction expenses, offering valuable insights into the real-world feasibility of this trading strategy.

In addition to the Distance Method (DM), cointegration tests are commonly employed in various alternative methods for mean reversion trading. \citeauthor{vidyamurthy2004pairs} (\citeyear{vidyamurthy2004pairs}) elucidate a cointegration framework for mean reversion trading, drawing inspiration from Engle and Granger's error correction model representation of cointegrated series as presented in the seminal work of \citeauthor{engle1987co} (\citeyear{engle1987co}). \citeauthor{galenko2012trading} (\citeyear{galenko2012trading}) explore an active ETF trading strategy built upon cointegrated time series. Their research leverages the concept of cointegration to develop trading strategies within the context of exchange-traded funds. \citeauthor{leung2019constructing} (\citeyear{leung2019constructing}) construct cointegrated portfolios of cryptocurrencies using the Engle-Granger two-step approach and the Johansen cointegration test, exemplifying the versatility of cointegration in portfolio construction across diverse asset classes. \citeauthor{huck2015pairs} (\citeyear{huck2015pairs}) undertake a comparative analysis of the performance of the DM method and cointegration-based approaches using the components of the S$\&$P 500 index. This study sheds light on the relative merits of these two methodologies in the context of mean reversion trading.
 
Another popular approach in mean reversion trading is the stochastic spread method, which characterizes the path behavior of the spread using a stochastic process exhibiting mean-reverting properties. In this methodology, the construction of spreads and the extraction of trading signals typically rely on the analysis of parameters within the underlying stochastic model. For instance, \citeauthor{elliott2005pairs} (\citeyear{elliott2005pairs}) introduce a mean-reverting Gaussian Markov chain model to describe the dynamics of spreads. They utilize this model's estimates in comparison with observed spread data to make informed trading decisions. Expanding on this framework, \citeauthor{do2006new} (\citeyear{do2006new}) delve deeper into the methodology proposed by \citeauthor{elliott2005pairs} and present a generalized stochastic residual spread method. This method is designed to model relative mispricing, offering a broader perspective on the stochastic spread approach in mean reversion trading. In their comprehensive work, \citeauthor{meanReversionBook} (\citeyear{meanReversionBook}) provide insights into optimal mean reversion trading strategies founded on various stochastic modeling approaches. Their research encompasses a range of models, including the Ornstein-Uhlenbeck (OU) model, Exponential OU Model, and CIR Model, elucidating how these different models can be leveraged to optimize mean reversion trading strategies. This work contributes to the broader understanding of mean reversion trading by exploring the effectiveness of diverse stochastic frameworks.

Furthermore, various alternative methods for mean reversion trading have emerged in recent years. Some of these methods leverage copulas (\citeauthor{liew2013pairs} (\citeyear{liew2013pairs}), \citeauthor{xie2016pairs} (\citeyear{xie2016pairs})), Principal Component Analysis (PCA) (\citeauthor{avellaneda2010statistical} (\citeyear{avellaneda2010statistical}), and machine learning techniques (\citeauthor{guijarro2021deep} (\citeyear{guijarro2021deep})) to identify trading opportunities based on statistical patterns and relationships among assets. In addition to these methodologies, new optimization algorithms have been proposed to construct spreads with maximum in-sample mean reversion. Notably, some of these algorithms are designed for automation, enabling the simultaneous analysis of a large number of stocks (\citeauthor{d2011identifying} (\citeyear{d2011identifying}), \citeauthor{LeungJizeSasha_Automatrica} (\citeyear{LeungJizeSasha_Automatrica})). These advancements reflect the ongoing innovation in the field of mean reversion trading, offering traders and researchers a diverse set of tools and techniques to explore.

In the domain of optimal timing strategies for trading mean-reverting spreads, several notable studies have contributed to the understanding of this field. \citeauthor{ekstrom2011optimal} (\citeyear{ekstrom2011optimal}) investigate the optimal single liquidation timing in the context of the Ornstein-Uhlenbeck (OU) model with zero long-run mean and no transaction costs. \citeauthor{song2009stochastic} (\citeyear{song2009stochastic}) introduce a numerical stochastic approximation scheme to solve for optimal buy-low-sell-high strategies over a finite time horizon. \citeauthor{leung2015optimal} (\citeyear{leung2015optimal}) address an optimal double stopping problem, providing optimal entry and exit decision rules. They derive analytical solutions for both the entry and exit problems under the assumption of an OU process. These studies collectively contribute to the advancement of optimal timing strategies for trading mean-reverting spreads, offering various perspectives and methodologies for tackling this challenging problem. The current paper is motivated by the model introduced by these prior researchers, proposing a sequential optimal stopping problem to determine optimal trading times without assuming specific dynamics for the spread.

The utilization of the signature method in the paper to address the optimal stopping problem aligns with recent developments in mathematical finance. \citeauthor{bayer2021optimal} (\citeyear{bayer2021optimal}) have introduced a novel approach to solving optimal stopping problems based on signature theory. Importantly, this method does not require specific assumptions about the underlying stochastic process other than it being a continuous (geometric) random rough path. The core concept revolves around considering classic stopping times approximated as signature optimal stopping times, which can be decided through a functional of the signature associated with the stochastic process. \citeauthor{bayer2021optimal} demonstrate that optimizing over these classes of signature stopping times effectively solves the original optimal stopping problem. This transformation allows for the problem to be reformulated as an optimization task dependent solely on the truncated signature. Furthermore, \citeauthor{bayer2021optimal} introduce a numerical solution to the signature optimal stopping problem by formulating the problem as an optimization task involving the minimization of a loss function associated with the truncated signature. This numerical approach provides a practical tool to implement the signature-based framework for solving optimal stopping problems, making it applicable to a wide range of real-world scenarios. They illustrate the practicality of this method in the context of optimal stopping for fractional Brownian motions.

Indeed, motivated by the innovative methodology introduced by \citeauthor{bayer2021optimal} (\citeyear{bayer2021optimal}), our paper embarks on an exploration of the potential application of the signature framework to tackle the optimal entry and exit problem in mean reversion trading. This extension highlights the remarkable versatility and promising capabilities of signature-based methods in effectively addressing a wide spectrum of challenges within financial contexts.

\section{Preliminaries of Signature}
\label{sec3}
Signature methods play an important role in this paper. In this section, we will introduce the theory of signatures that will serve as a foundational framework for the article.

\subsection{Tensor algebra}
A signature is an infinite series that takes values in a specific graded space known as the tensor algebra. To provide a solid foundation for understanding signatures, we will begin by introducing the fundamentals of tensor algebra.

The tensor product of two vectors $\mathbf{u} = (u_i)$ and $\mathbf{v} = (v_j)$ is given by:
$$
    (\mathbf{u}\otimes \mathbf{v})_{i,j} = u_i v_j
$$
For example, if $\mathbf{u} = (u_1, u_2) \in \mathbb{R}^2$ and $\mathbf{v} = (v_1, v_2, v_3) \in \mathbb{R}^3$, then their tensor product is given by
\[
\mathbf{u} \otimes \mathbf{v} = 
\begin{pmatrix}
u_1 v_1 & u_1 v_2 & u_1 v_3 \\
u_2 v_1 & u_2 v_2 & u_2 v_3 \\
\end{pmatrix} \in \mathbb{R}^{2 \times 3}.
\]
The concept of the tensor product can indeed be extended to accommodate multiple vectors or tensors. To illustrate, consider the example of taking the 3-way tensor product of three vectors $\mathbf{u} = [1,2],\ \mathbf{v}=[3,4],\ \mathbf{w}=[5,6]$:
$$
\begin{aligned}
&\begin{pmatrix}
1 \\ 2
\end{pmatrix} \otimes \begin{pmatrix}
3 \\ 4
\end{pmatrix} \otimes \begin{pmatrix}
5 \\ 6
\end{pmatrix} 
= \begin{pmatrix}
1 \cdot \begin{pmatrix}
3 \\ 4
\end{pmatrix} \otimes \begin{pmatrix}
5 \\ 6
\end{pmatrix} \\
2 \cdot \begin{pmatrix}
3 \\ 4
\end{pmatrix} \otimes \begin{pmatrix}
5 \\ 6
\end{pmatrix}
\end{pmatrix} 
= \begin{bmatrix}
\begin{array}{cc}
15 \ 18 \\ 20 \ 24 \\
\midrule[\arrayrulewidth]
30 \ 36 \\ 40 \ 48
\end{array}
\end{bmatrix}
\end{aligned}    
$$

In this context, the resulting tensor $\mathbf{u}\otimes \mathbf{v}\otimes \mathbf{w}$ is a $2\times 2\times 2$ array, where the $(i,j,k)$-th entry is the product of the $i$-th entry of $u$, the $j$-th entry of $v$, and the $k$-th entry of $w$.

Let \(U\) and \(V\) represent two vector spaces with bases \(\{\mathbf{u}_1,\ldots, \mathbf{u}_m\}\) and \(\{\mathbf{v}_1,\ldots,\mathbf{v}_n\}\), respectively. The tensor product space \(U\otimes V\) is defined as the vector space formed by the linear combinations of all possible products \(\mathbf{u}_i\otimes \mathbf{v}_j\), where \(i\) ranges from 1 to \(m\) and \(j\) ranges from 1 to \(n\). Formally, we express this as follows:
$$
U\otimes V = span\{\mathbf{u}_i\otimes \mathbf{v}_j | 1\leq i\leq m, 1\leq j\leq n\}.
$$
The tensor product space \(U\otimes V\) possesses a natural basis \(\{\mathbf{u}_i\otimes \mathbf{v}_j \,|\, 1\leq i\leq m, 1\leq j\leq n\}\). As an illustrative example, consider \(U=\mathbb{R}^2\) and \(V=\mathbb{R}^3\). In this case, the tensor product space \(U\otimes V\) has a dimension of \(2\times 3=6\), and a possible basis can be represented as follows:
$$
\begin{aligned}
\mathbf{u}_1 \otimes \mathbf{v}_1 &=
\begin{pmatrix}
1\\
0
\end{pmatrix}
\otimes
\begin{pmatrix}
1\\
0\\
0
\end{pmatrix}
=
\begin{pmatrix}
1 \ 0\\
0 \ 0\\
0 \ 0\\
\end{pmatrix}
\\
\mathbf{u}_1 \otimes \mathbf{v}_2 &=
\begin{pmatrix}
1\\
0
\end{pmatrix}
\otimes
\begin{pmatrix}
0\\
1\\
0
\end{pmatrix}
=
\begin{pmatrix}
0 \ 0\\
1 \ 0\\
0 \ 0\\
\end{pmatrix} \\
\vdots \\
\mathbf{u}_2 \otimes \mathbf{v}_3 &=
\begin{pmatrix}
0\\
1
\end{pmatrix}
\otimes
\begin{pmatrix}
0\\
0\\
1
\end{pmatrix}
=
\begin{pmatrix}
0 \ 0\\
0 \ 0\\
0 \ 1\\
\end{pmatrix}
\end{aligned}
$$
The concept of tensor product for two spaces can be readily extended to  multiple spaces. Let \(\mathbb{R}^d\), where \(d \geq 1\), be a finite-dimensional real vector space with a basis \(\{\mathbf{e}_1,\cdots,\mathbf{e}_d\}\). Then, \((\mathbb{R}^d)^{\otimes n}\) is the tensor product of \(n\) copies of \(\mathbb{R}^d\), resulting in a vector space with \(d^n\) dimensions. Each element in this tensor product space is represented as a tensor formed by taking the tensor product of \(n\) vectors, each belonging to \(\mathbb{R}^d\). We can define a basis for \((\mathbb{R}^d)^{\otimes n}\) as follows:
\[
\{\mathbf{e}_{i_1}\otimes \cdots \otimes \mathbf{e}_{i_n}\ |\ i_j\in\{1,\cdots,d\}\ \text{for}\ j=1,\cdots,n\}.
\]
Based on \((\mathbb{R}^d)^{\otimes n}\), we can provide the definition for the extended tensor algebra.

\begin{definition} (Extended tensor algebra) The extended tensor algebra over $\mathbb{R}^d$ is defined by 
$$
T((\mathbb{R}^d)) := \{ \textbf{a} = (a_n)_{n=0}^\infty\ |\ a_n\in (\mathbb{R}^d)^{\otimes n}\}.
$$
Given $\textbf{a} = (a_i)_{i=0}^\infty$, $\textbf{b} = (b_i)_{i=0}^\infty \in T((\mathbb{R}^d))$, define the sum and product by 
$$
\textbf{a} + \textbf{b} := (a_i + b_i)_{i=0}^\infty
$$
$$
\textbf{a} \otimes \textbf{b} :=\left(\sum_{k=0}^i a_k \otimes b_{i-k} \right)_{i=0}^\infty.
$$
We also define the action on $\mathbb{R}$ given by $\lambda \textbf{a} := (\lambda a_i)_{i=0}^\infty$ for $\lambda\in \mathbb{R}$.
\end{definition}

Likewise, we can define the tensor algebra as the space of all finite sequences of tensors. Additionally, we can define the truncated tensor algebra as the space of sequences of tensors of a specified, fixed length. These definitions allow us to work with sequences of tensors, accommodating various mathematical operations and analyses in this paper.

\begin{definition} (Tensor algebra) The tensor algebra over $\mathbb{R}^d$, denoted by $T(\mathbb{R}^d)\subset T((\mathbb{R}^d))$, is given by 
$$
T(\mathbb{R}^d) := \{ \textbf{a} = (a_n)_{n=0}^\infty\ |\ a_n \in (\mathbb{R}^d)^{\otimes n} \ \text{and}\ \exists N\in \mathbb{N}\ \text{such that}\ a_n=0\ \forall n\geq N \}.
$$
\end{definition}

\begin{definition} (Truncated tensor algebra)
The truncated tensor algebra of order $n\in \mathbb{N}$ over $\mathbb{R}^d$ is defined by
$$
T^{(N)}(\mathbb{R}^d) := \{ \textbf{a} = (a_n)_{n=0}^\infty\ |\ a_n \in (\mathbb{R}^d)^{\otimes n} \ \text{and}\ a_n=0 \ \forall n\geq N \}.
$$
\end{definition}

In the context of the dual space and linear functions on the tensor algebra, let \(\{\mathbf{e}_1^*,\cdots,\mathbf{e}^*_d\}\) be a dual basis for \((\mathbb{R}^d)^*\), where \((\mathbb{R}^d)^*\) represents the dual space of \(\mathbb{R}^d\). This dual space consists of all continuous linear functions mapping from \(\mathbb{R}^d\) to \(\mathbb{R}\). We can define a basis for \(((\mathbb{R}^d)^*)^{\otimes n}\) as follows:
\[
\{\mathbf{e}^*_{i_1}\otimes \cdots \otimes \mathbf{e}^*_{i_n}\ |\ i_j\in\{1,\cdots,d\}\ \text{for}\ j=1,\cdots,n\}.
\]
We denote the dual space of \(T((\mathbb{R}^d))\) as \(T((\mathbb{R}^d)^*) := ((\mathbb{R}^d)^*)^{\otimes n}\).

It's important to note that there exists a natural pairing between \(T((\mathbb{R}^d)^*)\) and \(T((\mathbb{R}^d))\), which we denote as \(\langle \cdot, \cdot \rangle: T((\mathbb{R}^d)^*) \times T((\mathbb{R}^d)) \rightarrow \mathbb{R}\). Consequently, the dual space \(T((\mathbb{R}^d)^*)\) encompasses all continuous linear functions defined on the extended tensor algebra \(T((\mathbb{R}^d))\).

\subsection{Signature}
We will now introduce a pivotal concept in this paper: the signature of a path. For a given path \(X:[0, T] \mapsto \mathbb{R}^d\), we denote the coordinate paths as \((X^1_t, \cdots, X^d_t)\), where each \(X^i:[0, T] \mapsto \mathbb{R}\) represents a real-valued path. For any single index \(i\in\{1, \cdots, d\}\) and \(0\leq a\leq b\leq T\), we define the quantity:
\[
S(X)^i_{a,b} = \int_{a<s<b}dX_s^i = X_b^i - X_a^i,
\]
which signifies the increment of the \(i\)-th coordinate of the path within the time interval \(t\in[a, b]\). 

For any pair of indices \(i,j \in \{1, \cdots, d\}\), we define the second-order path integral:
\[
S(X)^{i,j}_{a,b} = \int_{a<r<s<b} dX_r^i dX_s^j,
\]
and this concept can be generalized to arbitrary orders. For any integer \(n\geq1\) and a collection of indices \(i_1, \cdots, i_n \in \{1, \cdots, d\}\), we define:
\[
S(X)^{i_1,\cdots,i_n}_{a,b} = \int_{a<t_1<\cdots<t_n<b} dX_{t_1}^{i_1} \cdots dX_{t_n}^{i_n},
\]
referred to as the \(n\)-fold iterated integral of \(X\) along the indices \(i_1, \cdots, i_n\). The vector comprising all the \(n\)-fold iterated integrals of \(X\) is defined as:
\[
\mathbb{X}_{a,b}^n := \left(S(X)^{i_1,\cdots,i_n}_{a,b} \ \vline \ i_1, \cdots, i_n \in \{1, \cdots, d\}\right).
\]
In fact, \(\mathbb{X}_{a,b}^n\) belongs to \((\mathbb{R}^d)^{\otimes n}\), and we can express it concisely using the tensor product symbol:
\[
\mathbb{X}_{a,b}^n = \int_{a<t_1<\cdots<t_n<b} dX_{t_1} \otimes\cdots\otimes dX_{t_n} \in (\mathbb{R}^d)^{\otimes n}.
\]

\begin{example}
Consider a path $X:[0, T] \mapsto \mathbb{R}^3$, then 
$$
\mathbb{X}_{a,b}^1 = \left(S(X)^{1}_{a,b}, S(X)^{2}_{a,b}, S(X)^{3}_{a,b} \right),
$$
$$
\mathbb{X}_{a,b}^2 = \left(S(X)^{1,1}_{a,b}, S(X)^{1,2}_{a,b}, S(X)^{1,3}_{a,b}, S(X)^{2,1}_{a,b}, S(X)^{2,2}_{a,b}, S(X)^{2,3}_{a,b}, S(X)^{3,1}_{a,b}, S(X)^{3,2}_{a,b}, S(X)^{3,3}_{a,b} \right). 
$$
\end{example}

Now, we can introduce the definition of a signature.

\begin{definition} (Signature) Let $0 \leq a < b \leq T$. For a path $X:[0, T] \rightarrow \mathbb{R}^d$, we define the signature of $X$ over $[a, b]$ by
$$
\mathbb{X}_{a,b}^{<\infty} := (1, \mathbb{X}_{a,b}^1,\cdots, \mathbb{X}_{a,b}^n,\cdots) \in T((\mathbb{R}^d)),
$$
where
$$
\mathbb{X}_{a,b}^n := \int_{a<t_1<\cdots<t_n<b} dX_{t_1} \otimes\cdots\otimes dX_{t_n} \in (\mathbb{R}^d)^{\otimes n}.
$$
Similarly, the truncated signature of order $N$ is defined by
$$
\mathbb{X}_{a,b}^{\leq N} := (1, \mathbb{X}_{a,b}^1, \cdots,\mathbb{X}_{a,b}^N) \in T^{(N)}((\mathbb{R}^d)).
$$
\end{definition}

If we refer to the signature of \(X\) without specifying the interval over which the signature is taken, we will implicitly refer to \(\mathbb{X}_{0,T}^{<\infty}\).



In practical applications of the signature method, we typically transform the original path into an augmented path and then compute the truncated signature of the augmented path as the input features for any algorithm. Specifically, given a \(d\)-dimensional path \(X_t: [0,T] \rightarrow \mathbb{R}^d\), we denote its augmentation by \(\widehat{X}_t=(t, X_t)\in \mathbb{R}^{1+d}\), and define the augmented signature \(\widehat{\mathbb{X}}^{\leq \infty}\) and \(\widehat{\mathbb{X}}^{\leq \lfloor p\rfloor}\) as the signature and \(\lfloor p\rfloor\)-truncated signature of \(\widehat{X}_t\), respectively. 

Due to the first dimension of the augmented path (time) being monotonically increasing, the signature \(\widehat{\mathbb{X}}^{\leq \infty}\) provides a complete characterization of \(\widehat{X}\) (and therefore \(X\)). See \citeauthor{hambly2010uniqueness}(\citeyear{hambly2010uniqueness}). However, computing the entire signature with infinite length is not feasible, so the truncated signature with a fixed order \(p\) is used in the practice of the signature method. As a result, the time series to be analyzed in this paper will first be transformed to its truncated augmented signature for analysis.

\subsection{Signature Optimal Stopping Method}
\label{sec4}
In this section, we introduce the signature optimal stopping method proposed by \citeauthor{bayer2021optimal}(\citeyear{bayer2021optimal}), which will be applied in our approach to solve the proposed sequential signature optimal stopping problems.

\subsubsection{Signature Optimal Stopping Time}
The general optimal stopping problem involves determining an optimal stopping time that maximizes the value of the underlying process. Fix the probability space $(\Omega,\mathcal{F}, \mathbb{P})$. Suppose $X_t$ is a stochastic process and denote the filtration generated by $X$ as $(\mathcal{F}_t)$, $\mathcal{F}_t = \sigma(X_s: 0\leq s\leq t)$. Consider a payoff process $Y:[0,T]\times\Omega\rightarrow \mathbb{R}$ being a real-valued continuous stochastic process adapted to the filtration $(\mathcal{F}_t)$, then the optimal stopping problem for $Y_t$ is
\begin{equation} \label{eq1}
    \sup_{\tau\in\mathcal{T}} \mathbb{E}[Y_{\tau \wedge T}],
\end{equation}
where $\tau$ is a stopping time adapted to $(\mathcal{F}_t)$ and $\mathcal{T}$ is the space of all $(\mathcal{F}_t)$-stopping times

Similarly, the signature optimal stopping problem also involves determining an optimal stopping time to maximize the expected payoff of a stochastic process before a fixed time. However, the key difference between the two problems lies in the filtration used to define stopping times, which fundamentally affects the information available for making stopping decisions.
Specifically, let $(\Omega ,{\mathcal{F}}, \mathbb{P})$ be a probability space. Given a $d$-dimensional path $X_t: [0,T] \rightarrow \mathbb{R}^d$ and $\widehat{\mathbb{X}}$ denotes its augmented signature in $\widehat{\Omega}_T^p$. The filtration generated by the signatures are defined as $(\mathcal{G}_t)$, $\mathcal{G}_t = \sigma(\widehat{\mathbb{X}}_{0,s}:0\leq s\leq t)$ and denote by $\mathcal{S}$ the space of all $(\mathcal{G}_t)$-stopping times. $\mathcal{S}$ is referred as the space of signature stopping times. Consider a payoff process $Y:[0,T]\times\Omega\rightarrow \mathbb{R}$ being a real-valued continuous stochastic process adapted to the filtration $(\mathcal{G}_t)$, then the signature optimal stopping problem for $Y_t$ is:
\begin{equation} \label{eq2}
    \sup_{\tau\in\mathcal{S}} \mathbb{E}[Y_{\tau \wedge T}]
\end{equation}
where $\tau$ is a signature stopping time adapted to $(\mathcal{G}_t)$.

In summary, the general optimal stopping problem utilizes \((\mathcal{F}_t)\), a filtration based on the observable trajectory of the stochastic process \(X_t\), to determine stopping times \(\tau\) using data up to time \(t\). Conversely, the signature optimal stopping problem employs \((\mathcal{G}_t)\), generated from \(X_t\)'s augmented signatures \(\widehat{\mathbb{X}}\), which encapsulate both positions and dynamic patterns over time, offering richer information for stopping decisions. This enriched filtration potentially provides a more detailed insight into \(X_t\)'s behavior, guiding the derivation of stopping times.

\subsubsection{Randomized Stopping Time}
The fundamental concept for addressing the signature optimal stopping problem lies in the utilization of randomized stopping times. As demonstrated by \citeauthor{bayer2021optimal}, randomized stopping times are, in fact, equivalent to signature stopping times. Consequently, the signature optimal stopping problem, as depicted in Equation~(\ref{eq2}), can be effectively transformed into an equivalent problem known as the randomized optimal stopping problem.

Firstly, we introduce the concept of randomized stopping time. Define \(\widehat{\Omega}^p_T\) as the space of all the \(p\)-truncated augmented signatures, that is, elements of \(\widehat{\Omega}^p_T\) are \(p\)-truncated signatures of some augmented paths \((t, X_t)\). For a given time horizon $T>0$, we define $\Lambda_T := \bigcup_{t\in[0,T]} \widehat{\Omega}_t^p$, referring to it as the space of stopped signatures. Then, we denote a continuous stopping policy $\theta$ as a Borel measurable mapping $\theta: \Lambda_T \rightarrow \mathbb{R}$, and the space of continuous stopping policies as $\Theta:= C(\Lambda_T, \mathbb{R})$. $\theta$ is called a stopping policy because it can be employed to define the randomized stopping time based on its mapping value.

\begin{definition} (Randomized stopping time) Let $Z$
be a non-negative random variable independent of augmented signature $\widehat{\mathbb{X}}$
and such that $P(Z = 0) = 0$. For a continuous stopping policy $\theta\in\Theta$, we define the randomized stopping time by
\begin{equation} \label{eq3}
    \tau^r_\theta := \inf \left\{ t\geq 0: \int_0^{t\wedge T} \theta(\widehat{\mathbb{X}}|_{[0,s]})^2 ds \geq Z \right\}
\end{equation}
where $\inf\phi = +\infty$.
\end{definition}

Randomized stopping times, which are derived from continuous stopping policies, possess the capability to approximate signature stopping times, as demonstrated in the following theorem \cite[Proposition~4.2]{bayer2021optimal}.

\begin{theorem}\label{the4.2} For every stopping time $\tau\in\mathcal{S}$, there exists a sequence $\theta_n\in\Theta$ such that the randomized stopping times $\tau^r_{\theta_n}$ satisfy $\tau^r_{\theta_n}\rightarrow\tau$ almost surely as $n\rightarrow\infty$. In particular, if $\mathbb{E}[\|Y\|_\infty] < \infty$, then
\begin{equation} \label{eq4}
    \sup_{\theta\in\Theta} \mathbb{E}[Y_{\tau_\theta^r\wedge T}] = \sup_{\tau\in\mathcal{S}} \mathbb{E}[Y_{\tau\wedge T}].
\end{equation}
\end{theorem}

An astonishing result is that the general stopping time can be approximated by using stopping policies that are linear functions of the signature. The space of linear signature stopping policies $\Theta_{l} \subset \Theta$ is defined as
\[
    \Theta_{l} = \{ \theta\in\Theta: \exists l\in T((\mathbb{R}^{1+d})^*)\ \text{such that}\ \theta(\widehat{\mathbb{X}}|_{[0,t]}) = \langle l, \widehat{\mathbb{X}}_{0,t}^{<\infty} \rangle\ \forall \widehat{\mathbb{X}}|_{[0,t]} \in \Lambda_T \}.
\]
Note that every $l\in T((\mathbb{R}^{1+d})^*)$ defines a stopping policy $\theta_l \in\Theta$ by setting $\theta(\widehat{\mathbb{X}}|_{[0,t]}) = \langle l, \widehat{\mathbb{X}}_{0,t}^{<\infty} \rangle $. Then we introduce the following notation for randomized stopping times associated to linear signature stopping policies 
\begin{equation} \label{eq5}
    \tau^r_l := \tau^r_{\theta_l} =  \inf \left\{ t\geq 0: \int_0^{t\wedge T} \langle l, \widehat{\mathbb{X}}_{0,s}^{<\infty} \rangle^2 ds \geq Z \right\}.
\end{equation}

The subsequent theorem, detailed in \cite[Proposition~5.3]{bayer2021optimal}, demonstrates that the randomized stopping times linked to linear signature stopping policies are able to approximate general randomized stopping times.

\begin{theorem} \label{the4.3} Assume that $Z$ has a continuous density and $\mathbb{E}[\|Y\|_\infty]<\infty$. Then
\[
    \sup_{\theta\in\Theta} \mathbb{E}[Y_{\tau_\theta^r\wedge T}] = \sup_{\theta\in\Theta_{l}} \mathbb{E}[Y_{\tau_\theta^r\wedge T}].
\]
It follows that 
\begin{equation}\label{eq6}
    \sup_{\theta\in\Theta} \mathbb{E}[Y_{\tau_\theta^r\wedge T}] = \sup_{l\in T((\mathbb{R}^{1+d})^*)} \mathbb{E}[Y_{\tau_l^r\wedge T}].
\end{equation}
\end{theorem}

Continuing our exploration, we turn our attention to the evaluation of the optimal stopping problem with regard to a specific stopping policy \(\theta\). This evaluation plays a crucial role as it serves as a pivotal loss function, which will be utilized in the process of identifying the optimal linear policy during training. \cite[Proposition~4.4]{bayer2021optimal} offers comprehensive details into this matter. 

\begin{theorem}\label{the4.4} Let $F_Z$ denote the cumulative distribution function of $Z$. Then 
\begin{equation}\label{eq7}
    \mathbb{E}[Y_{\tau^r_\theta\wedge T}] = \int_0^T Y_t d\widetilde{F}(t) + Y_T(1 - \widetilde{F}(T)) = Y_0 + \int_0^T(1 - \widetilde{F}(t))dY_t
\end{equation}
where the second integral is implicitly defined by integration by parts and 
$$
\widetilde{F}(t) :=F_Z\left( \int_0^t\theta(\widehat{\mathbb{X}}|_{[0,s]})^2ds \right).
$$
In particular, if $Z$ has a density $f_Z$,
\begin{equation}\label{eq8}
\mathbb{E}[Y_{\tau^r_\theta\wedge T}] = \int_0^T Y_t\ \theta(\widehat{\mathbb{X}}|_{[0,t]})^2\ f_Z\left( \int_0^t\theta(\widehat{\mathbb{X}}|_{[0,s]})^2ds \right)dt + Y_T(1 - \widetilde{F}(T)).
\end{equation}
\end{theorem}

\subsubsection{Implementation of Signature Optimal Stopping Method}
In this section, we detail the implementation of solving the optimal signature stopping problem, utilizing the previously introduced theorems.

The initial step involves discretizing the paths, where we establish a time grid comprising equidistant points denoted as $\{t_j\}_{j=0,1,\ldots,n}$, with $t_j = \frac{j}{n}T$. After discretization, we compute the signature of the time-augmented path \((t_j, X_{t_j})_{0\leq j \leq n}\), denoted by \(\widehat{\mathbb{X}}_{0, t_j}\).
Then, we adapt the definition of randomized stopping times to the discrete-time setting. Given a positive random variable $Z$ and a continuous stopping rule $\theta\in\Theta$, we define the discrete version of the randomized stopping time in (\ref{eq3}) as follows:

\begin{equation}\label{eq9}
    \nu^r_\theta := \inf \left\{ 0\leq j\leq n: \sum_{i=0}^{j} \theta(\widehat{\mathbb{X}}|_{[0,t_i]})^2 \geq Z \right\}
\end{equation}

The discrete version of randomized stopping times associated with linear signature
stopping policies is defined as:
\begin{equation}\label{eq10}
    \nu^r_l := \inf \left\{ 0\leq j\leq n:  \sum_{i=0}^{j} 
    \langle l, \widehat{\mathbb{X}}^{<\infty}_{0, t_i}\rangle^2 \geq Z \right\}
\end{equation}

The similar results established in Theorems \ref{the4.2} to \ref{the4.4} also extend to discrete stopping times. Building upon Theorem \ref{the4.4}, we can represent the expected payoff in (\ref{eq7}) as follows:

\begin{equation}\label{eq11}
\mathbb{E}[Y_{\nu^r_\theta}] = Y_0 + \sum_{j=0}^{n-1} G_Z\left(\sum_{i=0}^{j} \theta(\widehat{\mathbb{X}}|_{[0,t_i]})^2\right) (Y_{j+1} - Y_j),
\end{equation}
where $G_Z = 1 - F_Z$. 

We can further elaborate on the loss function for linear stopping policies based on the regular form of the expected payoff provided above. Given a data-set comprising $M$ samples in the format $(X_j^{(m)}, Y_j^{(m)})_{1\leq m\leq M, 0\leq j\leq n}$ and a predetermined truncation level $N\geq 1$, we compute the $N$-truncated augmented signature of the path on each sub-interval $[0, t_j]$ for each series of samples, define as
$\widehat{\mathbb{X}}_{0, t_j}^{(m)}$. Subsequently, we establish the following loss function for the linear stopping policy $\theta_l$ based on Equation (\ref{eq11}):

\begin{equation}\label{eq12}
\text{loss}(l) = - \frac{1}{M}\sum_{i=1}^M\left\{Y_0^{(m)} + \sum_{j=0}^{n-1} G_Z \left(\sum_{i=0}^j \langle l, \widehat{\mathbb{X}}_{i}^{(m)}\rangle^2\right) (Y^{(m)}_{j+1} - Y^{(m)}_j) \right\}.
\end{equation}

Equations \eqref{eq10} and \eqref{eq12} will serve as our foundational framework for addressing the signature optimal stopping problem. Initially, we utilize Equation \eqref{eq12} to derive the optimal linear policy \(l^*\) from the training samples. Subsequently, we determine the optimal stopping time for the target time series by applying Equation \eqref{eq10}. 

\section{Signature Optimal Mean Reversion Trading} 
\label{sec5}
The main contribution of this paper is the development of an innovative trading strategy tailored for mean reversion markets. Our methodology hinges on the formulation of a sequence of optimal stopping problems, aimed at identifying the most advantageous moments for both market entry and exit.

In addition, we propose a solution to these optimal stopping problems through a sequential signature-based optimal stopping framework. This serves as a robust approximation to the broader, more general optimal stopping challenges commonly found in financial mathematics. Notably, our approach is highly adaptable, accommodating any form of mean-reverting dynamics. This framework empowers us to derive a sequence of optimal timing solutions for trading activities, thereby elevating the existing mean-reversion trading rules in both theoretical constructs and real-world applications.

\subsection{Sequential Signature Optimal Stopping Framework}
\label{sec4.1}
Given that a price process or portfolio value following a mean-reverting process, our primary goal is to investigate the optimal timing for opening a long position and subsequently closing the position. To achieve this, we first introduce a framework that involves solving a sequence of signature optimal stopping problems.

\subsubsection{Long Strategy}
Suppose that the investor trades on a spread or portfolio whose value process $(X_t)_{t\geq 0}$ follows a mean reverting process. Recall that $\widehat{\mathbb{X}}$ denotes the augmented signature of $X_t$ and the filtration generated by signatures is $(\mathcal{G}_t)$, where $\mathcal{G}_t = \sigma(\widehat{\mathbb{X}}_{0,s}:0\leq s\leq t)$.  Denote by $\mathcal{S}$ the space of all $(\mathcal{G}_t)$-stopping times. 

Let us define the shorthand notation for the expectation conditional on the starting value $x$ as $E_x\{\cdot\} := E\{\cdot|X_0=x\}$. If the long position is opened at some time $\tau_0>0$, then the investor will pay the value $X_{\tau_0}$ plus a constant transaction cost $c > 0$. Assuming that $r > 0$ is the investor’s subjective constant discount rate, the total cost of entering a long position is $e^{-r\tau_0}(X_{\tau_0} + c)$. Our goal is to minimize the expectation of the entering cost. Therefore, we obtain the following optimal stopping problem for finding the optimal time of entering a long position:
\begin{equation}
\label{eq13}
\sup_{\tau_0\in \mathcal{S}} E_{x_0} \{e^{-r\tau_0}(- X_{\tau_0} - c)\},
\end{equation}
where $\mathcal{S}$ is the space of all $(\mathcal{F}_t)$-stopping times. In a word, we are minimizing the expected cost of entering a long position, given the initial point $x_0$.

After obtaining the optimal time to enter a long position, denoted by $\tau_0^*$, the next question is when to exit the position. If the position is closed at time $\nu_0$, the investor will receive the value $X_{\nu_0}$ and pay a constant transaction cost $\hat{c} > 0$. To maximize the expected discounted value, the investor solves the optimal stopping problem:
\begin{equation}
\label{eq14}
\sup_{\nu_0\in \mathcal{S}} E_{x_{\tau^*_0}} \{e^{-\hat{r}\nu_0}(X_{\nu_0} - \hat{c})\}.
\end{equation}
Here, we aim to maximize the expectation of discounted earnings conditional on the starting point $x_{\tau^*_0}$. Note that the pre-entry and post-entry discount rates, $\hat{r}$ and $r$, as well as the entry and exit transaction costs $\hat{c}$ and $c$, can differ in our analysis.

Then we can solve a sequence of optimal stopping problems iteratively to obtain the optimal entry and exit times. That is, for $i = 1, 2, 3, \ldots$, we solve the following equations iteratively:
\begin{equation} 
\label{eq15}
\begin{aligned}
\sup_{\tau_i \in \mathcal{S}} E_{x_{\nu_{i-1}^*}} &\{e^{-r\tau_i}(- X_{\tau_i} - c)\}, \\
\sup_{\nu_i \in \mathcal{S}} E_{x_{\tau^*_{i}}} &\{e^{-\hat{r}\nu_i}(X_{\nu_i} - \hat{c})\}. 
\end{aligned}
\end{equation}
Thus, we get the optimal times of trading $\{\tau^*_0, \nu^*_0, \tau^*_1, \nu^*_1, \tau^*_2, \nu^*_2, \cdots \}$. 

\subsubsection{Short Strategy}
In the short strategy, our main objective is to determine the optimal timing for opening and subsequently closing a short position, similar to the long strategy. We only need to change the order of equations for the optimal stopping problems in (\ref{eq14}). Specifically, we obtain the first optimal entry time $\nu_0^*$ and exit time $\tau_0^*$ by solving:
\begin{equation}
\label{eq16}
\begin{aligned}
&\sup_{\nu_0\in \mathcal{S}} E_{x_{0}} \{e^{-r\nu_0}(X_{\nu_0} - c)\}, \\
&\sup_{\tau_0\in \mathcal{S}} E_{x_{\nu_0^*}} \{e^{- \hat{r} \tau_0}(- X_{\tau_0} - \hat{c})\}.
\end{aligned}
\end{equation}
Next, we solve the following equations iteratively for $i = 1, 2, 3, \cdots$:
\begin{equation}
\label{eq17}
\begin{aligned}
\sup_{\nu_i \in \mathcal{S}} E_{x_{\tau_{i-1}^*}} &\{e^{-r\nu_i}( X_{\nu_i} - c)\}, \\
\sup_{\tau_i \in \mathcal{S}} E_{x_{\nu^*_{i}}} &\{e^{-\hat{r}\tau_i}(- X_{\tau_i} - \hat{c})\}
\end{aligned}
\end{equation}
to obtain the optimal trading times $\{\nu^*_0, \tau^*_0, \nu^*_1, \tau^*_1, \nu^*_2, \tau^*_2, \cdots \}$. The intuition behind is opening a short position when the spread price reaches a high value, and closing the position as the price drops significantly.

\subsection{Non-Randomization of Signature Stopping Times}
\label{sec4.2}
Following the signature optimal stopping framework, we attempt to deploy linear signature stopping times, denoted \(\nu^r_l\), as specified in Equation \eqref{eq10}, to ascertain the optimal stopping times for a given stochastic process. The optimization of the linear policy \(l^*\) is conducted via a minimization of the loss function presented in Equation \eqref{eq12}, and subsequently, the policy is applied to determine the signature optimal stopping time for the observed time series. However, the randomization introduced by the random variable \(Z\) in Equation \eqref{eq10} typically precludes a deterministic solution of stopping time. To address this challenge, we propose a novel approach by replacing the random variable \(Z\) with a constant \(k\) to provide a deterministic solution for the optimal stopping time, while  preserving the applicability of Equations \eqref{eq10} and \eqref{eq12}.
 
Specifically, the non-randomized stopping time associated with linear signature stopping policies is then defined as follows:
\begin{equation}\label{eq18}
    \nu_l := \inf \left\{ 0 \leq j \leq n:  \sum_{i=0}^{j} \langle l, \widehat{\mathbb{X}}^{<\infty}_{0, t_i}\rangle^2 \geq k \right\},
\end{equation}
where \(k\) is a positive constant. This approach allows us to compute the stopping time without relying on randomness, offering a deterministic solution. Therefore, in our practical implementation, we calculate the optimal stopping time based on Equation (\ref{eq18}) once we have determined the optimal linear stopping policy \(l^*\).

\begin{figure}
    \centering
    \includegraphics[width=10 cm]{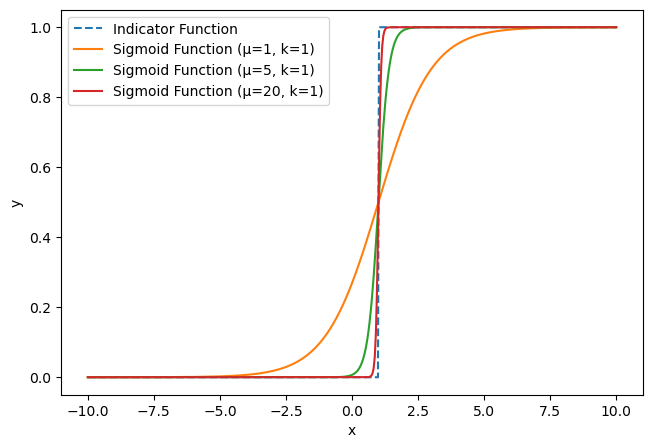}
    \caption{Approximating indicator function using sigmoid function.}
    \label{fig1}
\end{figure}

Substituting the random variable \(Z\) with a constant \(k\) transforms the cumulative distribution function of \(Z\) into an indicator function. Thus, this substitution introduces non-differentiability into the loss function. Such non-differentiability poses computational challenges, particularly in the application of gradient-based optimization methods which are foundational to our algorithm's practical implementation. To overcome this challenge, we propose a refinement to the loss function as originally defined in Equation (\ref{eq12}). Initially, we substitute \(G_Z\) with \(G_k\), defined as \(G_k = 1 - F_k\). Here, \(F_k(x) = \mathbb{I}\{x > k\}\), with \(\mathbb{I}\) as the indicator function. To facilitate gradient-based optimization, we approximate \(F_k\) using a sigmoid function centered at \(k\). The approximation \(\widehat{F}_k(x)\) is given by:
\[
\widehat{F}_k(x) = \frac{1}{1 + e^{-\mu \cdot (x - k)}}
\]
where \(\mu\) is a large positive constant that sharpens the sigmoid function. Figure \ref{fig1} illustrates the approximation of the indicator function using a sigmoid function, providing a visual representation of how the sigmoid curve serves as a smoothed substitute for the step-like behavior of the indicator function. In the subsequent experiments, we employ a \(\mu\) value of 20.

Subsequently, we employ the following modified loss function in our practical implementation:
\begin{equation}\label{eq19}
\text{loss}(l) = - \frac{1}{M}\sum_{i=1}^M \left\{ Y_0^{(m)} + \sum_{j=0}^{n-1} \widehat{G}_k \left( \sum_{i=0}^j \langle l, \widehat{\mathbb{X}}_{i}^{(m)} \rangle^2 \right) (Y^{(m)}_{j+1} - Y^{(m)}_j) \right\},
\end{equation}
where \(\widehat{G}_k = 1 - \widehat{F}_k\). This formulation allows for gradient-based methods to be effectively employed for optimization.

In this section, we have provided a comprehensive overview of our implementation process of signature optimal stopping method, with a particular emphasis on the non-randomization of signature stopping times and the refined loss function. This loss function in \eqref{eq19} serves as a crucial component in our training procedure. Upon successfully obtaining the optimal linear policy, we proceed to compute the corresponding optimal signature stopping time using Equation \eqref{eq18}. This comprehensive methodology help us solve the complex sequential signature optimal stopping problems in section \ref{sec4.1}.

\subsection{Training Samples Generation}
To implement the tailored signature optimal stopping method introduced in Section~\ref{sec4.2}, this section delves into the training procedures employed in our methodology. The whole process can be applied to both long and short strategies, but we will focus on the long strategy in this discussion. 

Recall that the long strategy framework is grounded in the formulations presented by equations~\eqref{eq13}, \eqref{eq14}, and \eqref{eq15}. To solve the first signature optimal stopping problem (\ref{eq13}), we need to obtain the optimal linear function $l_0^*$ by minimizing the loss function given in equation (\ref{eq19}). However, we can only observe one path $X_t, 0\leq t\leq T$, which is actually our target time series. So we first generate training samples of $X_t$ starting at a given initial value $x$, and the corresponding payoff process $Y_t = e^{-rt}(- X_{t} - c)$. We then use these samples to minimize the loss function to obtain the optimal linear function $l_{0,1}^*$. Once we obtain the optimal linear stopping policy $l_{0,1}^*$, we can use it to approximate the optimal entry time $\tau_0^*$ by the linear non-randomized optimal stopping time defined in equation (\ref{eq18}). Specifically, the optimal entry time is given by:
$$
    \tau_0^* = \inf \left\{ 0\leq j\leq n: \sum_{i=0}^{j}
    \langle l_{0,1}^*, \widehat{\mathbb{X}}_{0,t_i}^{\leq N} \rangle^2
     \geq k \right\}
$$
where $n$ is the number of discretization intervals, $\widehat{\mathbb{X}}_{0,t_i}^{\leq N}$ is the $N$-truncated augmented signature of the path $X_t$ from time $0$ to time $t_i$, and $k$ is a positive threshold value.

To address the optimal exit problem specified in Equation \eqref{eq14},  it is necessary to simulate sample paths of $X_t$ starting from the new initial value $X_{\tau_0^*}$ over the time interval $[\tau_0^*, T]$. The payoff process is calculated as $Y_t = e^{-\hat{r}t}(X_t - \hat{c})$. We then can obtain another optimal linear function $l_{0,2}^*$ using the generated samples, and the optimal exit time is given by:
$$
    \nu_0^* = \inf \left\{ 0\leq j\leq n: \sum_{i=0}^{j}
    \langle l_{0,2}^*, \widehat{\mathbb{X}}_{\tau_0^*,\tau_0^*+t_i}^{\leq N} \rangle^2
     \geq k \right\}
$$
where $\widehat{\mathbb{X}}_{\tau_0^*,\tau_0^*+t_i}^{\leq N}$ is the $N$-truncated augmented signature of the path of $X_t$ from time $\tau_0^*$ to time $\tau_0^*+t_i$. We can continue this computation process to solve the sequence of problems in (\ref{eq15}) and thus obtain $\{\tau^*_0, \nu^*_0, \tau^*_1, \nu^*_1, \tau^*_2, \nu^*_2, \cdots \}$, where each pair of values $(\tau^*_n, \nu^*_n)$ corresponds to the optimal entry and exit times for the $n$-th trading. To be more specific, for $n\geq 1$, we have:
$$
\begin{aligned}
&\text{Generate sample paths of } X_t \text{ starting from } X_{\nu_{n-1}^*} \text{ over } [\nu_{n-1}^*, T] \text{ and } Y_t = e^{-rt}(-X_t - c); \\
&\text{Use the signature optimal stopping method to obtain } l^*_{{n},1} \text{ and } \tau_{n}^*; \\
&\text{Generate sample paths of } X_t \text{ starting from } X_{\tau_{n}^*} \text{ over } [\tau_{n}^*, T] \text{ and } Y_t = e^{-\hat{r}t}(X_t - \hat{c}); \\
&\text{Use the signature optimal stopping method to obtain } l^*_{{n},2} \text{ and } \nu_{n}^*.
\end{aligned}
$$
We continue this process until we reach the end of the trading period. At each step, we obtain the optimal linear function and the corresponding optimal stopping time, which can be used to guide our trading decisions.

As for the methods for generating training sample paths from the observed time series $X_t$, we have several choices. The initial approach is block bootstrapping \cite{berkowitz2000recent}, a widely utilized method in the bootstrapping of time series data. This technique involves resampling contiguous blocks of data points, thereby preserving the data's inherent structures. An alternative strategy involves fitting the observed path to a parametric statistical or stochastic model, subsequently generating new samples from this model. For instance, the path could be modeled using an Autoregressive (AR) model or an Ornstein-Uhlenbeck (OU) process. Additionally, generative machine learning approaches, such as time-series Generative Adversarial Networks (GAN) \cite{yoon2019time}, present a modern method for sample path generation. Researchers have the flexibility to select the most suitable method to fulfill the requirements of the algorithm, considering the specific characteristics of the observed data.

\section{Numerical Experiments}
\label{sec6}
In this section, we undertake numerical experiments to rigorously assess the efficacy of our proposed methodologies. We initiate our evaluation by scrutinizing the performance of the proposed optimal trading timing approach through its application to simulated mean reversion spreads. Following this, we extend our analysis to real-world scenarios by evaluating the mean reversion trading performance of our method across four carefully chosen, highly correlated asset pairs in the U.S. stock markets.

\subsection{Experiments on Simulated Data}
In this section, we conduct simulated experiments on the effectiveness of the proposed signature optimal mean reversion trading method.

\subsubsection{Test on Signature Optimal Stopping Method}
In this part, we conduct simulations to gauge the effectiveness of the signature-based optimal stopping method in the context of mean reversion spreads. We employ the Ornstein-Uhlenbeck (OU) process to simulate a mean-reverting spread, governed by the following stochastic differential equation:
\[
    dX_t = \theta (\mu - X_t) dt + \sigma dW_t.
\]
Our primary aim is to identify the supremum of the expected value of the OU process at the signature optimal stopping time, formally expressed as:
\[
\sup_{\tau \in \mathcal{S}} \mathbb{E}[X_{\tau \wedge T}].
\]
By leveraging this simulation framework, we intend to rigorously evaluate the capability of the signature optimal stopping method to accurately identify optimal stopping values in mean-reverting contexts.

In applying the signature optimal stopping method to solve the optimal stopping problem for an Ornstein-Uhlenbeck (OU) process, our approach consists of two distinct phases. In the first phase, we synthesize a dataset consisting of 100 training samples. These samples are used to compute the optimal linear function \(l^*\) that minimizes the loss function, as delineated in Equation (\ref{eq19}). In the second phase, we generate a supplementary set of 10 testing samples. Utilizing the previously computed optimal linear function \(l^*\), we ascertain the optimal stopping times for these test cases in accordance with Equation (\ref{eq18}) and record the corresponding values of path at that time. This two-step procedure enables a comprehensive evaluation of the method's performance, both in terms of its optimization capabilities and its generalizability to new instances.

\begin{figure}
    \centering
    \includegraphics[width=8 cm]{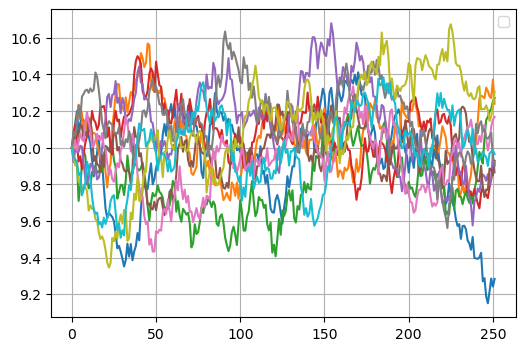}
    \caption{The paths of the 10 testing samples.}
    \label{fig2}
\end{figure}

In this specific experiment, the parameters that define the Ornstein-Uhlenbeck (OU) process are set as follows: \(\mu = 10\), \(\theta = 10\), and \(\sigma = 1\). As depicted in Figure~\ref{fig2}, the graph displays the trajectories of the 10 testing samples. The calculated average optimal stopping value is found to be 10.23. Importantly, the maxima for the majority of these paths are observed to cluster around 10.4, which lends credence to the computed optimal stopping value of 10.23. This result compellingly validates the effectiveness of the  signature optimal stopping methodology, as it consistently produces high optimal stopping values for the simulated OU processes.

We have also executed a series of tests to examine the optimal stopping values generated by the signature method across a range of parameter combinations within the Ornstein-Uhlenbeck (OU) process. The aggregate findings of these tests are consolidated in Table~\ref{tab1}. Significantly, all the optimal stopping values obtained exceeded the long-term mean of 10. A particularly insightful observation is the positive correlation between the optimal stopping value and the volatility parameter \(\sigma\). This observed relationship underscores the sensitivity of the optimal stopping decision to the volatility of the underlying process, highlighting the nuanced interplay between process parameters and stopping outcomes.

\begin{table}[H]
    \centering
    \begin{tabular}{ c c c|c } 
    \hline\hline
    $\mu$ & $\sigma$ & $\theta$ & Optimal Stopping Value \\ 
    \hline
    1 & 1.0 & 10 & 10.1284  \\
    5 & 1.0 & 10 & 10.2311  \\
    10 & 1.0 & 10 & 10.2332  \\
    15 & 1.0 & 10 & 10.2005  \\
    20 & 1.0 & 10 & 10.1703  \\
    \hline
    10 & 0.1 & 10 & 10.0025  \\
    10 & 0.5 & 10 & 10.0713  \\
    10 & 1.0 & 10 & 10.2264  \\
    10 & 1.5 & 10 & 10.3599 \\
    10 & 2.0 & 10 & 10.5173 \\
    \hline\hline
    \end{tabular}
    \caption{The optimal stopping value estimated by signature method on simulated mean-reversion process. Parameters $\mu \in \{1, 5, 10, 15, 20\}$ and  $\sigma \in \{0.1, 0.5, 1.0, 1.5, 2.0\}$.}
    \label{tab1} 
\end{table}

\subsubsection{Simulation on Signature Optimal Mean Reversion Trading}
In this section, we conduct simulations using the Ornstein-Uhlenbeck (OU) process to test the effectiveness of our proposed optimal trading timing method for identifying both entry and exit times in mean-reverting markets. We continue to simulate the mean-reversion paths by an OU model. This simulation study serves to validate the utility of our approach in generating actionable trading decisions based on the underlying stochastic process.

\begin{figure}[H]
    \subfigure[]{\includegraphics[width=6.5 cm]{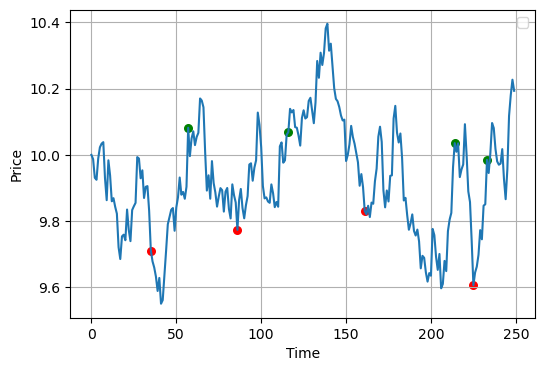}}
    \subfigure[]{\includegraphics[width=6.5 cm]{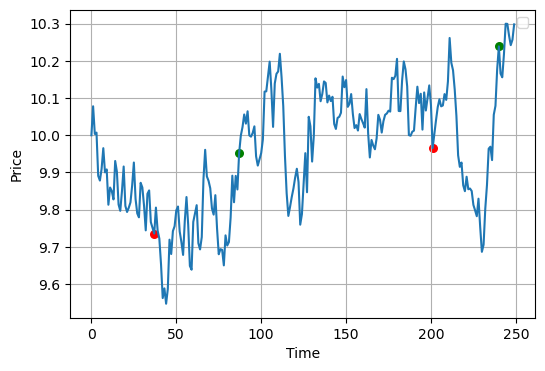}}\\
    \subfigure[]{\includegraphics[width=6.5 cm]{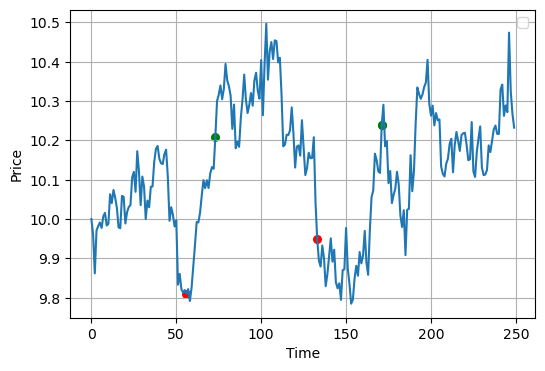}}
    \subfigure[]{\includegraphics[width=6.5 cm]{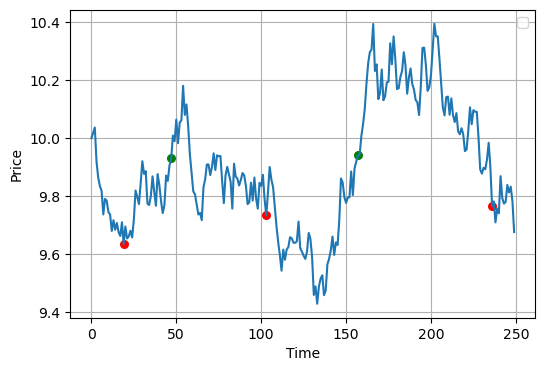}}
    \caption{The sequential optimal entry and exit times for a simulated mean reversion process based on optimal trade timing framework.}
    \label{fig3}
\end{figure}

Figure~\ref{fig3} displays the performance of our simulation, where we have marked optimal entry times using red points and optimal exit times with green points. This color-coded representation provides an intuitive snapshot of how adeptly our method identifies strategic moments to both enter and exit trades in a mean-reverting market. As is evident from the graph, our method proves highly effective, consistently point times that align well with the natural peaks and troughs of the mean-reverting process. These results lend robust empirical support to the theoretical analysis of our proposed trading timing methodology. Moreover, they offer compelling evidence that our approach can serve as a valuable tool for traders and portfolio managers seeking to exploit mean-reversion opportunities in financial markets. The simulation outcomes underscore the practical applicability and accuracy of our algorithm, confirming its utility for making well-informed, timely trading decisions based on the dynamics of the underlying stochastic process.

\subsection{Experiments in Real Market}
In this section, we broaden the application of our signature optimal mean reversion trading methodology to a real-world trading context. It is important to emphasize that our method operates without making any assumptions about the dynamics of the underlying traded spreads in real market.

Initially, we handpick four highly co-integrated stock pairs from diverse sectors within the U.S. market to create their corresponding spreads: WM-RSG from waste management, UAL-DAL from the airline industry, V-MA from financial services, and GS-MS from investment banking. A comprehensive description of these companies is provided in Table~\ref{tab2}. Data pertaining to the daily closing prices for these stocks was collected for the period spanning January 1, 2021, to December 31, 2022. The data was sourced via the Yahoo! Finance API\footnote{https://pypi.org/project/yfinance/}. 

\begin{figure}[H]
    \includegraphics[width=14 cm]{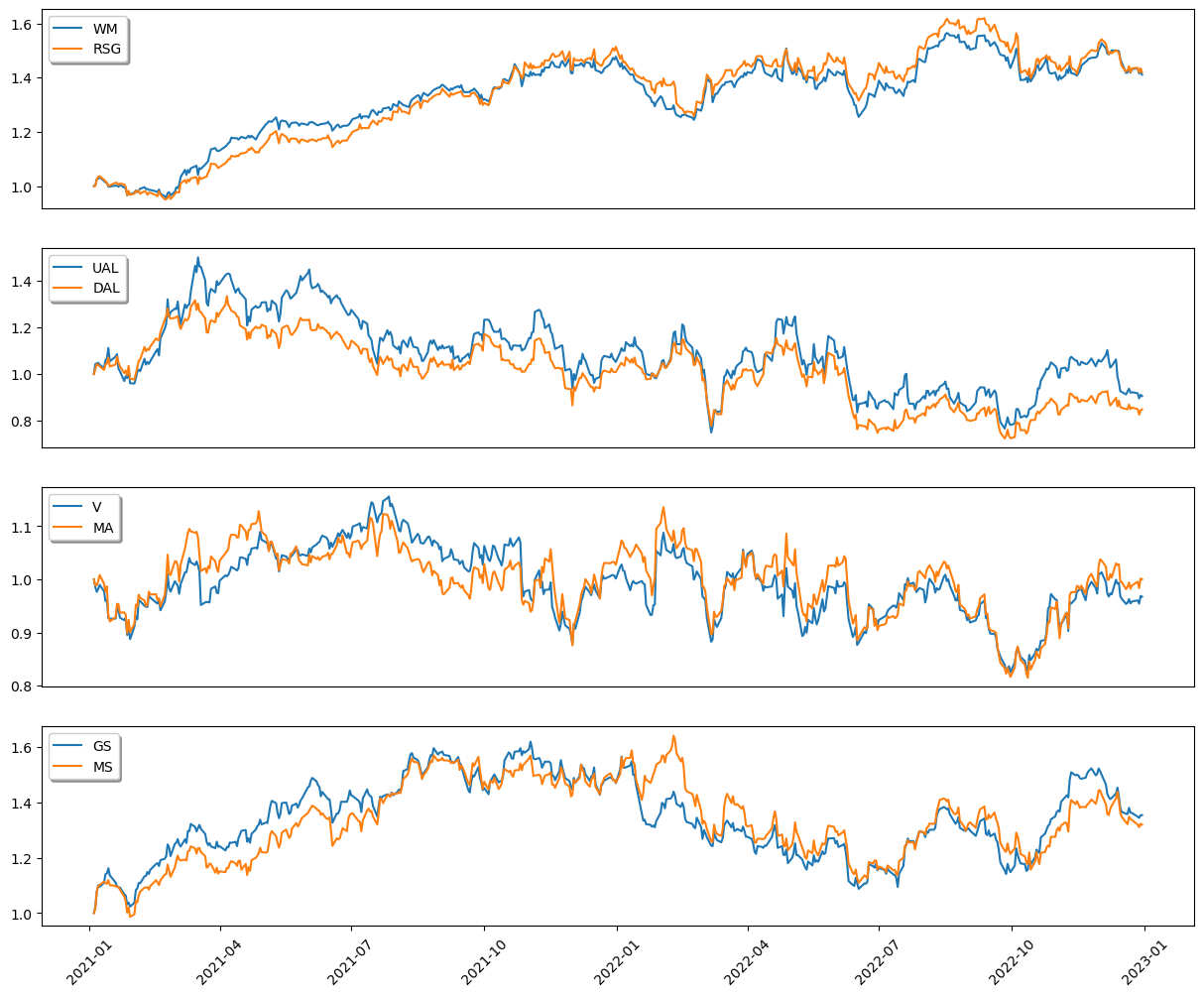}
    \caption{Normalized historical daily close prices of six pairs of stocks from 1/1/2021--12/31/2022. From top to bottom: WM-RSG, UAL-DAL, V-MA, GS-MS, where each pair shows visibly persistent comovement.}
    \label{fig4}
\end{figure}

Figure~\ref{fig4} depicts the price trajectories of each stock pair, revealing a notable similarity in the price paths within each pair. For the sake of clarity and comparability, we have normalized each price series by dividing it by its respective initial value. This normalization allows for a more straightforward visual assessment of the relative movements between the stocks in each pair, thereby setting the stage for the forthcoming trading experiments.

\begin{table}[H] 
    \centering
    \begin{tabular}{ c|c } 
    \hline\hline
    \textbf{Pairs Symbols}	& \textbf{Description}\\
    \hline 
    WM-RSG & Waste Management and Republic Services provide waste management \\
    & and environmental services \\
    \hline
    UAL-DAL & United Airlines and Delta Air Lines are two major American airline \\
    & operating a large domestic and international route network. \\ 
    \hline
    V-MA  & Visa and Mastercard are two large-cap stocks in the payment industry  \\
    \hline
    GS-MS & Goldman Sachs and Morgan Stanley are American multinational investment bank \\
    &and financial services company. \\
    \hline\hline
    \end{tabular}
    \caption{The stock pairs used for testing and the description of the company.}
    \label{tab2} 
\end{table}

\begin{figure}[H]
    \subfigure[WM-RSG]{\includegraphics[width=7 cm]{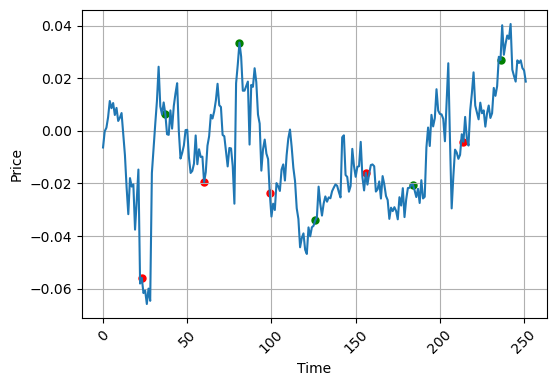}}
    \subfigure[UAL-DAL]{\includegraphics[width=7 cm]{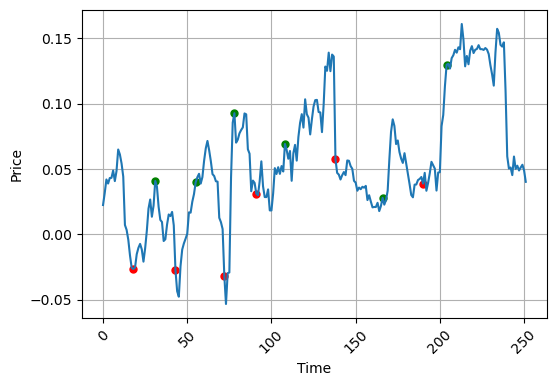}}\\
    \subfigure[V-MA]{\includegraphics[width=7 cm]{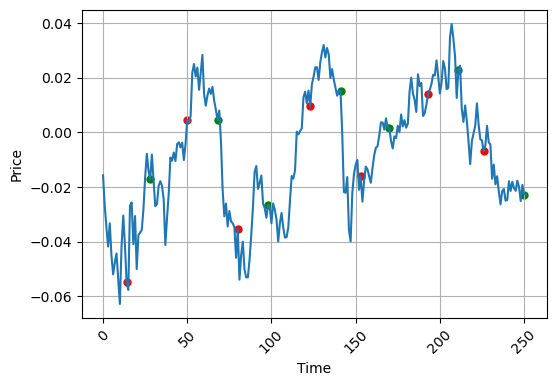}}
    \subfigure[GS-MS]{\includegraphics[width=7 cm]{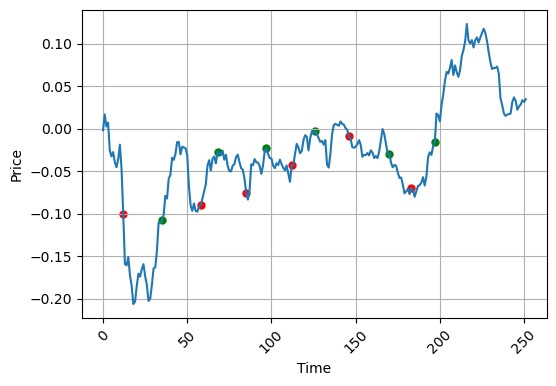}}
    \caption{The constructed spreads and optimal entry and exit time points. Red spot is entry point and Green spot is exit point.}
    \label{fig5}
\end{figure}

In the experiments, we simulated one full year of trading activity, as visually presented in Figure~\ref{fig5}. Specifically, we created asset pairs based on the daily adjusted closing prices from the first year, covering a period of 252 trading days. These pairs were then actively traded over the course of the subsequent year. By doing so, we aim to evaluate the real-world applicability and effectiveness of our signature-based optimal stopping method within a time frame that captures a meaningful range of market conditions. This setup allows us to assess how well the trading rules, derived from the initial year's data, generalize to unseen data in the following year. In the setup of the proposed signature optimal mean reversion trading method, the training samples generation procedure employed OU process and the threshold value $k =0.05$.

As illustrated in Figure~\ref{fig5}, our algorithm excels in identifying the opportunities of trading, consistently entering positions at low prices and exiting at higher prices. This behavior serves as a compelling demonstration to the efficacy of our signature-based optimal trading strategy. This strong performance underscores the algorithm's utility as a valuable tool for market participants seeking to exploit mean-reversion opportunities with optimized entry and exit timings.

In the final phase of our analysis, we contrast the performance of our signature-based optimal trading strategy with a baseline strategy across all four asset pairs. The baseline strategy follows a simple rule: go long on \(X_t\) when its price falls below \(MA(X_t) - k \times Std(X_t)\), and liquidate the position when the price rises above \(MA(X_t) + k \times Std(X_t)\). Here we write   $MA(X_t)$ and $Std(X_t)$ to represent  the moving average and standard deviation of past spread, respectively. 

\begin{figure}[H]
    \subfigure[WM-RSG]{\includegraphics[width=7 cm]{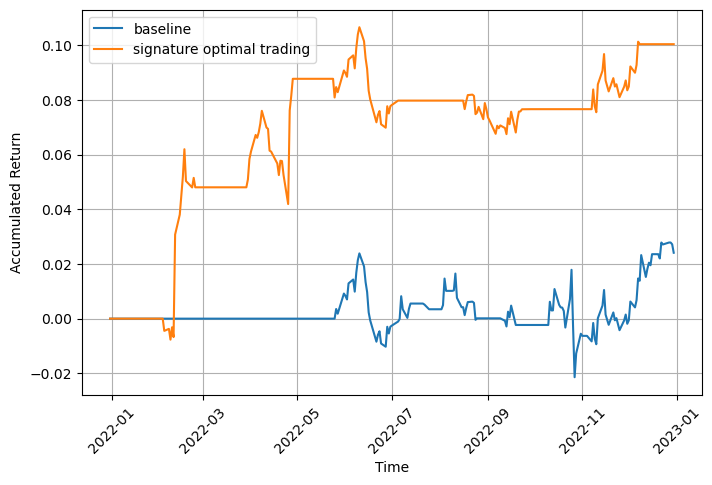}}
    \subfigure[UAL-DAL]{\includegraphics[width=7 cm]{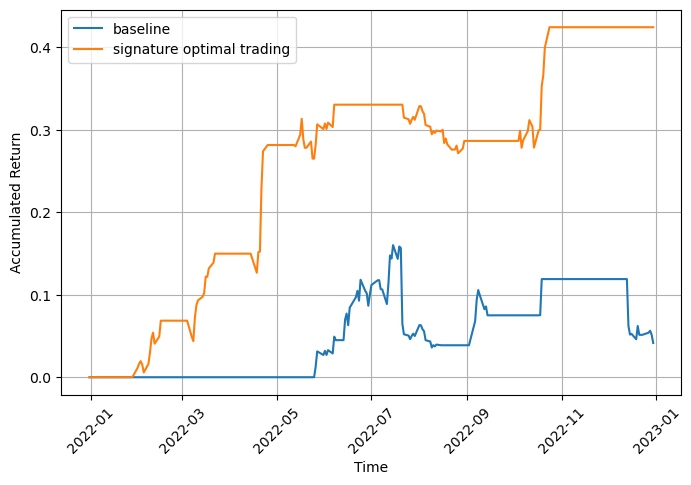}}\\
    \subfigure[V-MA]{\includegraphics[width=7 cm]{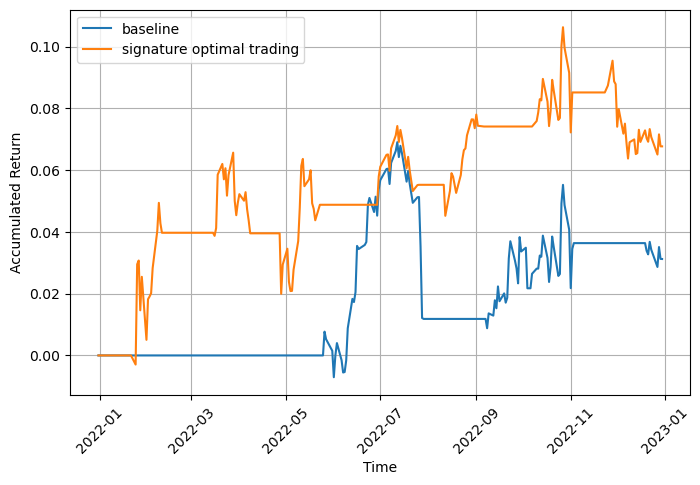}}
    \subfigure[GS-MS]{\includegraphics[width=7 cm]{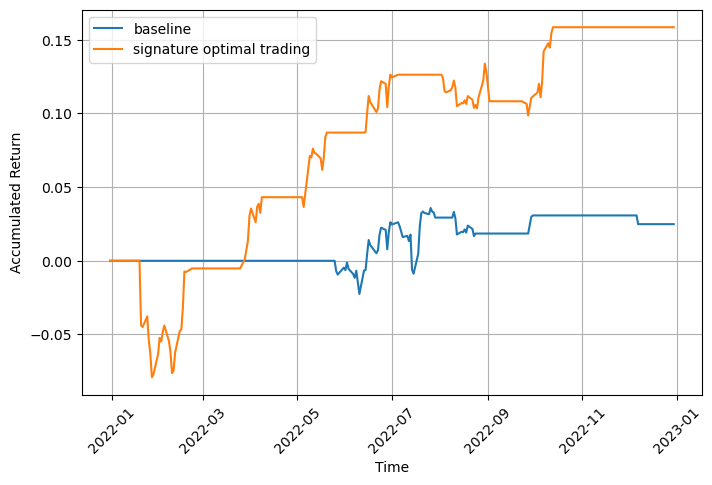}}
    \caption{The accumulative returns for baseline and SOT. }
    \label{fig6}
\end{figure}

Table~\ref{tab3} presents a thorough comparison of performance metrics for both the baseline and the signature optimal trading (SOT) strategies. In this evaluation, the baseline strategy employs a constant \(k\) value of 0.1 for its trading rules and leverages the preceding 100 samples to compute both the moving average and standard deviation. Figure~\ref{fig6} graphically illustrates the cumulative returns generated by each strategy over a one-year time frame. 

As revealed by Table~\ref{tab3}, our proposed signature optimal trading strategy substantially outperforms the baseline model in key performance metrics, including higher cumulative returns and a more favorable daily Sharpe ratio. This head-to-head comparison is specifically tailored to underscore the superior efficacy of the SOT approach. It compellingly attests to the enhanced performance and incremental value that our method brings to the table in trading scenarios. 

The visual representation of cumulative returns in Figure~\ref{fig6} serves as another pivotal element in our comparative assessment. It provides an intuitive snapshot of how the respective trading strategy compound over a one-year period. By offering this view, we further reinforce the idea that the SOT strategy is not only theoretically robust but also practically effective and adaptable to real-world market conditions. This comprehensive evaluation aims to highlight the enduring viability and real-world utility of the SOT methodology.

\begin{table}
    \begin{tabular}{ c | cc | cc | cc | cc }
        \hline\hline
	& \multicolumn{2}{c|}{\textbf{WM/RSG}} & \multicolumn{2}{c|}{\textbf{UAL/DAL}} & \multicolumn{2}{c|}{\textbf{V/MA}} & \multicolumn{2}{c}{\textbf{GS/MS}} \\ 
        & Baseline & SOT & Baseline & SOT & Baseline & SOT & Baseline & SOT \\ 
        \hline
        DailyRet ($\%$) & 0.0110 & 0.0390 & 0.0401 & 0.1443 & 0.0107 & 0.0275 & 0.0119 &  0.0604   \\ 
        DailyStd ($\%$) & 0.3513 & 0.4321 & 0.8244 & 0.8122 & 0.4177 & 0.5306 & 0.3362 &  0.5986 \\ 
        Sharpe	        & 0.0313 & 0.0903 & 0.0487 & 0.1777 & 0.0255 & 0.0519 & 0.0354 &  0.1012  \\
        MaxDD ($\%$)    & -1.9309 & -0.7674 & 0 & 0 & -0.7057 & -0.2954 & -2.2522 &  -7.9006  \\
        CumPnL ($\%$)   & 2.6363 & 10.0345 & 9.6431 & 42.4559 & 2.4857 & 6.7732 & 2.8829 & 15.8486   \\
        TradeNum        & 9 & 5 & 4 & 6 & 3 & 7 & 4 & 6 \\ 
	\hline\hline
    \end{tabular}
    \caption{Performance summary for all 4 pairs portfolios for baseline and the signature optimal trading (SOT). From the top to bottom rows, we report the annualized return, annualized standard deviation, Sharpe ratio, maximum drawdown and number of tradings.}
    \label{tab3}
\end{table}

\section{Conclusions}
\label{sec7}

In this research paper, we have presented an innovative approach to identifying optimal timing strategies for trading price spreads with mean-reverting properties. Specifically, we formulated a sequential optimal stopping problem that accounts for the timing of both position entry and liquidation, while also incorporating transaction costs. To tackle this challenging problem, we employed the signature optimal stopping method, a powerful tool for determining the optimal entry and exit times that maximize trading returns.

Our approach is distinguished by its versatility: it is designed to operate without any assumptions of mean reversion dynamics, making it broadly applicable to a variety of trading scenarios. To validate the efficacy of our methodology, we carried out a comprehensive set of numerical experiments. These experiments served to compare the performance of our approach against conventional mean-reversion trading rules. The results consistently demonstrated the superior performance of our signature-based optimal trading strategy in terms of key metrics such as cumulative returns and Sharpe ratios.

In summary, this paper makes a significant contribution to the field of quantitative finance by introducing a robust and adaptable method for optimizing trading decisions in mean-reverting markets. Our findings affirm that the signature optimal stopping method offers not only theoretical rigor but also practical utility, presenting a compelling solution to a complex problem in finance. While our focus has been on specific asset pairs and trading conditions, the general principles and techniques introduced here have the potential for broader applications, opening up avenues for future research and practical implementations.
 
\bibliographystyle{plainnat}
\bibliography{main}
 
\end{document}